\documentclass[twocolumn,showpacs,preprintnumbers]{revtex4}
\usepackage{graphicx}
\usepackage{dcolumn}
\usepackage{bm}

\input{tcilatex}

\begin{document}

\title{Classical Correlation in Quantum Dialogue}
\author{Yong-gang Tan$^{1,2}$ and Qing-yu Cai$^{1}$}
\email{qycai@wipm.ac.cn}
\affiliation{$^{1}$State Key Laboratory of Magnetic Resonances and Atomic and Molecular
Physics, Wuhan Institute of Physics and Mathematics, the Chinese Academy of
Sciences, Wuhan 430071, People's Republic of China}
\affiliation{$^{2}$Graduate School of the Chinese Academy of Sciences, Beijing 100049,
People's Republic of China}

\begin{abstract}
Classical communications are used in the post-processing procedure of
quantum key distribution. Since the security of quantum key distribution is
based on the principles of quantum mechanics, intuitively the secret key can
only be derived from the quantum states. We find that classical
communications are incorrectly used in the so-called quantum dialogue type
protocols. In these protocols, public communications are used to transmit
secret messages. Our calculations show that half of Alice's and Bob's secret
message is leaked through classical channel. By applying Holevo bound, we
can see that the quantum efficiency claimed in the quantum dialogue type of
protocols is not achievable.

\end{abstract}

\maketitle

\section{Introduction}

Quantum key distribution (QKD) is an unconditionally secure method by which
a private key can be created between two parties, Alice and Bob, who share a
quantum channel and a public authenticated\ classical channel. Since the
pioneer QKD protocol was presented by Bennett and Brassard in 1984~\cite%
{ben84}, its security has been studied intensively~\cite{may01,lo99,shor00}.
In BB84 protocol, Alice randomly selects one of four states in two
complementary bases to encode her secret message and Bob also randomly
selects one of the two bases to decode Alice's key bits. Consequently, basis
reconciliation is necessary in BB84 protocol. Recently, the quantum secure
direct communication (QSDC) protocols have been presented \cite%
{bos03,cai04,ca04,den04,deng04,lu05}. In the QSDC protocols, Bob can decode
Alice's encoded message directly after his measurement and they don't need
to do basis reconciliation. Based on the idea of QSDC, a new type of quantum
communication protocol, called quantum dialogue (QD), has been presented 
\cite{deg04, ngu04, ji06, yan07}.

In the entanglement-based QD protocol \cite{ngu04}, it is claimed that both
Alice and Bob can encode two-bit secret message on an EPR pair. After a
public announcement, both Alice and Bob can obtain two-bit secret message
from each other. That is, Alice and Bob can encode four-bit secret message
in one EPR pair. Likewise, in a single-photon QD protocol \cite{ji06}, both
Alice and Bob can encode one-bit secret message on a photon. So, both Alice
and Bob can obtain one-bit secret message from each other. It is claimed
that one qubit can be used to transmit two-bit secret message \cite{ji06},
where the quantum efficiency is four times than BB84.

In this paper, we prove that QD protocol presented by \cite{deg04, ngu04,
ji06, yan07} are insecure because classical communication is erroneously
used in these protocols. Alice and Bob's encoding operations are correlated
given the published measurement results. Our calculations show that Eve can
gain half information of Alice and Bob's secret message only by listening
the classical channel. By applying Holevo bound, we can see that the quantum
efficiency claimed in the quantum dialogue type of protocols \cite{deg04,
ngu04, ji06, yan07} is not achievable.

\section{review the quantum dialogue protocol}

There are three types of QD protocols, quantum dense key distribution using
entanglement \cite{deg04} which is the prototype of quantum dialogue,~QD
protocol based on EPR pairs \cite{ngu04} and QD protocol with single photon
sources~\cite{ji06}. Let us briefly review the idea of QD here \cite{ngu04}.
Alice first prepares an EPR pair in the singlet state $|\Psi_{AB}^{-}\rangle$%
, where $|\Psi_{AB}^{-}\rangle=\frac{1}{\sqrt{2}}(|0_{A}1_{B}%
\rangle-|1_{A}0_{B}\rangle)$. She keeps one qubit A in her laboratory and
sends the other qubit B to Bob. After receiving qubit B, Bob may randomly
select message mode (MM) or control mode (CM). In CM, Bob performs a local
measurement on qubit B and tells measurement results to Alice. After
receiving Bob's announcement, Alice also switches to CM and measures her
qubit A. Alice and Bob can estimate the fidelity of EPR pair after enough
runs of CM. The QD transmission will be aborted if the fidelity of the EPR
pair is lower than some certain threshold. In MM, after receiving qubit B,
Bob performs a unitary operation $U^{B}$ on qubit B to encode his secret
message and then sends it back to Alice. After receiving the back qubit B,
Alice first performs a unitary operation $U^{A}$ on qubit A and the state
becomes $|\Psi_{AB}\rangle=U^{A}U^{B}|\Psi_{AB}^{-}\rangle$. Next, Alice
announces her measurement result $|\Psi_{AB}\rangle$ through the classical
channel. Since Alice knows her own encoding operation $U^{A}$ and the
measurement result $|\Psi_{AB}\rangle$, she can exactly know Bob's encoding
operation $U^{B}$ to attain Bob's encoded message. Likewise, Bob can obtain
Alice's encoded message according to $|\Psi_{AB}\rangle$ and $U^{B}$.
Consequently, it seems as if both Alice and Bob can transmit\ one-bit
\textquotedblleft secret\textquotedblright\ message to each other
simultaneously so that we call this protocol QD. In Ref.\cite{ngu04}, both
Alice and Bob use four unitary operations\ $\sigma_{00}$, $\sigma_{01}$, $%
\sigma_{10}$, $\sigma_{11}$ to encode their secret message, 00, 01, 10, 11,
respectively. Consequently, both Alice and Bob can transmit two-bit secret
message to each other in each MM.

Likewise, in\ quantum dense key distribution protocol \cite{deg04}, Alice
first prepares one EPR pair in $|\Psi_{AB}^{-}\rangle$ and sends one qubit
to Bob. In MM, both Alice and Bob encode one-bit secret message and then
Alice announces her measurement results through classical channel.\ In this
way, they can transmit one-bit secret message to each other. Similarly, in
the single photon QD protocol\ \cite{ji06}, Alice first prepares a photon in
one of four states in two complementary bases and then both Alice and Bob
encode one-bit secret message on it. After Alice's announcement, both can
obtain one-bit secret message from each other.

\section{half secret message leaked through public channel}

At the first glance, the QD protocol is secure since no one except Alice and
Bob knows Alice or Bob's secret encoding operations to gain their secret
message. However, we will show in the following, anyone who can access Alice
and Bob's classical channel can gain half information about their secret
message. Eve's mean information gain on Alice and Bob's bits, $I(AB:E)$,
equals their relative entropy decrease~\cite{gin02}: 
\[
I(AB:E)=H_{a\ priori}-H_{a\ posteriori}, 
\]
where $H_{a\ priori}$ is the $a\ priori$ entropy and $H_{a\ postoriori}$ is
the $a\ postoriori$ entropy. In~Ref.\cite{ngu04}, the $a\ priori$ entropy
Alice and Bob shared are $4$ bits, that is, $H_{a\ priori}=4$. And the $a\
posteriori$ entropy of Eve is averaged over all possible results $r$. So she
can get $H_{a\ posterior}=\sum_{r}P(r)H(i|r)$, where $H(i|r)=-\sum
_{i}P(i|r)log_{2}[P(i|r)]$ is the conditional information entropy. By
listening the classical channel, Eve can obtain Alice's measurements results 
$|\Psi_{AB}^{-}\rangle$, $|\Psi_{AB}^{+}\rangle$, $|\Phi_{AB}^{-}\rangle$
and $|\Phi_{AB}^{+}\rangle$, i.e., $r=4$. Each of the four measurement
results corresponds to four of Alice's and Bob's operations $%
\sigma_{ij}^{A}\sigma_{kl}^{B}$. The true value table of their encoding
operations is presented in Table I. 
\begin{table}[ptb]
\caption{True value table of Alice's and Bob's encoding operations. $%
|\Psi_{AB}^{-}\rangle$, $|\Psi_{AB}^{+}\rangle$, $|\Phi_{AB}^{-}\rangle$, $%
|\Phi_{AB}^{+}\rangle$ are the final states of their encoding operations $%
\protect\sigma_{ij}^{A}\protect\sigma_{kl}^{B}$ on $|\Psi_{AB}^{-}\rangle$
and $i$, $j$, $k$, $l$ $\in$ 0,1.}%
\begin{ruledtabular}
\begin{tabular}
[c]{rccl}
$|\Psi_{AB}^{-}\rangle$ & $|\Psi_{AB}^{+}\rangle$ & $|\Phi_{AB}^{-}\rangle$ &
$|\Phi_{AB}^{+}\rangle$\\\hline
$\sigma_{00}^{A}\sigma_{00}^{B}$ & $\sigma_{11}^{A}\sigma_{00}^{B}$ &
$\sigma_{01}^{A}\sigma_{00}^{B}$ & $\sigma_{10}^{A}\sigma_{00}^{B}$\\
$\sigma_{01}^{A}\sigma_{01}^{B}$ & $\sigma_{10}^{A}\sigma_{01}^{B}$ &
$\sigma_{00}^{A}\sigma_{01}^{B}$ & $\sigma_{11}^{A}\sigma_{01}^{B}$\\
$\sigma_{10}^{A}\sigma_{10}^{B}$ & $\sigma_{01}^{A}\sigma_{10}^{B}$ &
$\sigma_{11}^{A}\sigma_{10}^{B}$ & $\sigma_{00}^{A}\sigma_{10}^{B}$\\
$\sigma_{11}^{A}\sigma_{11}^{B}$ & $\sigma_{00}^{A}\sigma_{11}^{B}$ &
$\sigma_{10}^{A}\sigma_{11}^{B}$ & $\sigma_{01}^{A}\sigma_{11}^{B}$\\
\end{tabular}
\end{ruledtabular}
\end{table}
And the final state, $|\Psi_{AB}^{-}\rangle$, $|\Psi_{AB}^{+}\rangle$, $%
|\Phi_{AB}^{-}\rangle$, or $|\Phi_{AB}^{+}\rangle$, would be published
through classical channel. For instance, after Alice announces her
measurement result $|\Psi_{AB}^{-}\rangle$, Eve may have that $P(\sigma
_{00}^{A}\sigma_{00}^{B}||\Psi_{AB}^{-}\rangle)=P(\sigma_{01}^{A}\sigma
_{01}^{B}||\Psi_{AB}^{-}\rangle)=P(\sigma_{10}^{A}\sigma_{10}^{B}||\Psi
_{AB}^{-}\rangle)=P(\sigma_{11}^{A}\sigma_{11}^{B}||\Psi_{AB}^{-}\rangle)=%
\frac{1}{4}$ (Here $P(x|y)$ is the probability of $x$ conditioned on $y$,
and we assume that Alice and Bob's operations are random). In this case,
Eve's information about Alice and Bob's secret message is $I(AB:E)=2$. That
is, Eve can obtain half information about Alice and Bob's secret message
only by listening the classical cannel.

Likewise, Eve can gain one-bit secret information only by listening the
classical channel in the quantum dense key distribution protocol \cite{deg04}%
. In \cite{ji06}, Eve can also obtain one-bit secret message by listening
the classical channel. Therefore, QD protocols are insecure even if Alice
and Bob hold a perfect quantum channel since half of secret message would be
leaked through the classical channel. On the other hand, we will show in the
following, the quantum efficiency claimed in QD violates Holevo bound.

\section{Violation of Holevo bound}

In QKD, a secret key is encoded in quantum states. So, the maximal secret
information that can be transmitted in each run is completely determined by
the quantum channel capacity. The capacity of a quantum channel is bounded
by the Holevo bound. If information is encoded on a state $\rho$, the
accessible information of $\rho$ is bounded by the Holevo quantity~\cite%
{hol73} 
\[
\chi(\rho)=S(\rho)-\sum_{i}p_{i}S(\rho_{i}). 
\]
where $\rho=\sum_{i}p_{i}\rho_{i}$. In quantum communication, the mutual
information $I(A:B)$ between Alice and Bob should be less than the Holevo
bound~\cite{cab00}, i.e., $I(A:B)\leq S(\rho)-\sum_{i}p_{i}S(\rho_{i})$. In 
\cite{ngu04}, the quantum channel capacity is bounded by $\log_{2}4=2$,
i.e., at most two-bit secret message can be encoded in each MM. As has been
discussed above, Alice and Bob encodes four bits secret message in each MM,
so that two-bit secret message will be leaked through classical channel. In
fact, Alice's and Bob's encoding operations are correlated by the formula 
\[
\sigma_{ij}^{A}\sigma_{kl}^{B}|\Psi_{AB}^{-}\rangle\rightarrow|\Psi_{(i%
\oplus j),(k\oplus l)}^{AB}\rangle. 
\]
So, Eve can directly obtain the correlation between Alice's and Bob's
encoding operations after the announcement of the final states. Let us
assume that Alice publishes her measurement result $|\Psi_{AB}^{-}\rangle$.
As discussed above, the possible operations of Alice's and Bob's are $%
\sigma_{00}^{A}$$\sigma_{00}^{B}$, $\sigma_{01}^{A}$$\sigma_{01}^{B}$, $%
\sigma_{10}^{A}$$\sigma_{10}^{B}$ and $\sigma_{11}^{A}$$\sigma_{11}^{B}$
(also see Table I) and then Eve can obtain partial information of Alice and
Bob's secret message.

\section{discussion and conclusion}

Although QD is insecure, it can still be applied to QKD as an approach to
generate raw key. If Alice and Bob have realized that half of their secret
message has leaked through classical channel, they can implement privacy
amplification to distill secure final key bits. Let us emphasize that
quantum efficiency can truly be improved by some other approaches. Note that
the efficient BB84 has already been proposed and its unconditional security
proof has also been presented \cite{lc05}.

In summery, QD is insecure because of the erroneous use of classical
communication which reveals classical correlations between Alice and Bob's
encoding operations. The classical channel is public so that everyone
including Eve can access it to attain the correlations between Alice and
Bob's encoding operations. Our calculations showed that half of Alice and
Bob's secret message would be leaked through classical channel in QD
protocols. The quantum efficiency claimed in QD violates the Holevo bound.

We note that when this study was completed, we found that the erroneous use
of classical communication in QD was independently pointed out by Gao 
\textit{et al}. \cite{gao08}.

\begin{acknowledgments}
This work is funded by Natural Science Foundation of China (Grant No.
10504039) and Chenguang Youth Project of Wuhan City.
\end{acknowledgments}

\end{document}